\documentclass[aps,prl,groupedaddress,twocolumn,10pt,reprint,superscriptaddress, longbibliography,floatfix]{revtex4-2}

\usepackage{amsmath}
\usepackage{amssymb}
\usepackage{amsthm}
\usepackage{bbm}
\usepackage{graphicx}
\usepackage{amsmath}
\usepackage{graphicx}
\usepackage{color}
\usepackage{upgreek}
\usepackage[linkcolor = blue, citecolor = blue, urlcolor = blue, colorlinks = true]{hyperref}
\usepackage{bm}
\usepackage[capitalise]{cleveref}
\usepackage{textcomp}
\usepackage{hyperref}
\usepackage[usenames,dvipsnames]{xcolor}
\usepackage[normalem]{ulem}
\usepackage{wasysym,siunitx}

\usepackage{siunitx}

\newcommand{\dd}[1]{\mathrm{d}#1}
\newcommand{\eps}{\epsilon}
\newcommand{\xm}{\bm{x}}

\newcommand{\xim}{\bm{\xi}}
\newcommand{\uhat}{\bm{\hat{u}}}
\newcommand{\J}{\bm{J}}
\newcommand{\Ptot}{P^{(0)}_{\mathrm{tot}}}
\newcommand{\Deff}{D_{\mathrm{eff}}}
\newcommand{\Dliq}{D_{\mathrm{liq}}}
\newcommand{\phiv}{\phi_{\mathrm{void}}}

\newcommand{\kesc}{k_{\mathrm{esc}}}
\newcommand{\ktrap}{k_{\mathrm{trap}}}

\renewcommand{\imath}[0]{\mathsf{i}}

\hypersetup{colorlinks=true, linkcolor=blue!50!black, urlcolor=blue!50!black, citecolor=blue!50!black}

\definecolor{ABpurple}{RGB}{128, 0, 128}
\definecolor{ABred}{RGB}{255, 0, 0}
\definecolor{ABgreen}{RGB}{0, 255, 0}
\definecolor{ABbrown}{RGB}{128, 64, 0}
\definecolor{ABblue}{RGB}{0, 0, 255}

\begin{document}
\title{Bacterial diffusion in disordered media, by forgetting the media} 
\author{Henry H. Mattingly}
\date{\today}
\email{hmattingly@flatironinstitute.org}
\affiliation{Center for Computational Biology, Flatiron Institute, Simons Foundation}

\begin{abstract}
We study bacterial diffusion in disordered porous media. Interactions with obstacles, at unknown locations, make this problem challenging. We approach it by abstracting the environment to cell states with memoryless transitions. With this, we derive an effective diffusivity that agrees well with simulations in explicit geometries. The diffusivity is non-monotonic, and we solve the optimal run length. We also find a rescaling that causes all of the theory and simulations to collapse. Our results indicate that a small set of microscopic features captures bacterial diffusion in disordered media.
\end{abstract}
\maketitle

\textit{Introduction ---} Transport of bacteria through spatially-structured environments is important for many natural and engineering processes. For example, bacterial transport through the body is important for the initiation and spread of infections \cite{siitonen_bacterial_1992, chaban_flagellum_2015, ribet_how_2015, matilla_effect_2018, keegstra_ecological_2022}; in the Earth's subsurface, it impacts contamination of water sources and bioremediation efforts \cite{ginn_processes_2002, simon_1_2002, adadevoh_chemotaxis_2016, adadevoh_chemotaxis_2018}; and in soils, it affects plant-microbe interactions that are relevant to agricultural yields \cite{scharf_chemotaxis_2016, keegstra_ecological_2022, pantigoso_rhizosphere_2022, chepsergon_rhizosphere_2023}. Building models of bacterial transport in these environments is essential for understanding and controlling the dynamics of these processes. Furthermore, bacteria have evolved diverse motility strategies \cite{grognot_more_2021}, which might are likely adaptive to the environments they naturally encounter. 

We focus on diffusive transport of bacteria through disordered porous media. Modeling bacterial motility in these environments is difficult because their active self-propulsion causes complex interactions with surfaces. Unlike passive particles \cite{berezhkovskii_diffusivity_2003,berezhkovskii_effective_2003}, actively-driven swimmers tend to follow and accumulate on walls \cite{galajda_wall_2007, elgeti_wall_2013, elgeti_microswimmers_2016, bechinger_active_2016}. Recent experiments tracking bacteria swimming through jammed hydrogels have found that cells ``hop" through void spaces and get ``trapped" in dead ends \cite{bhattacharjee_bacterial_2019, bhattacharjee_confinement_2019}. Furthermore, hydrodynamic interactions can cause swimmers to align with walls, swim in circles, or scatter from obstacle surfaces \cite{diluzio_escherichia_2005, berke_hydrodynamic_2008, drescher_fluid_2011, takagi_hydrodynamic_2014, spagnolie_geometric_2015, lauga_bacterial_2016, bianchi_holographic_2017, makarchuk_enhanced_2019, dehkharghani_self-transport_2023}. To make matters worse, we usually don't know the geometry of the environment that the cell navigates.

This problem has long been an active area of research. One approach is to start from a diffusion equation with walls as boundary conditions, and then derive the model's long-time behavior \cite{valdes-parada_upscaling_2009}. However, this cannot the capture interactions with obstacle surfaces. Licata et al. derived a diffusion coefficient using a model of a run-and-tumble particle \cite{berg_chemotaxis_1972,berg_e_2004}, in which long runs caused the cell to become immobilized until its next tumble \cite{licata_diffusion_2016}. But this model cannot be quantitatively compared to systems with explicit geometries. Rigorous results have been derived using lattice models, but these models cannot capture the effects of different wall shapes and are difficult to compare to experiments \cite{bertrand_optimized_2018, rizkallah_microscopic_2022}. Beautiful results have also been derived using periodically-spaced obstacles, but they relied on precise knowledge of the environment geometry \cite{alonso-matilla_transport_2019,jakuszeit_diffusion_2019}. Recent work extracted hop and trap statistics from simulations with explicit obstacles, and then used these statistiscs to predict the diffusion coefficient \cite{kurzthaler_geometric_2021}. But this approach precludes \textit{a priori} prediction of the diffusion coefficient, or expressing it in terms of cell and environmental parameters. Also recently, a trajectory-based theory derived a diffusivity using the probabilities of various tumble and encounter events \cite{saintillan_dispersion_2023}. However, this theory is somewhat limited to low obstacle density and short runs. A diffusivity that captures experimental measurements was recently reported \cite{dehkharghani_self-transport_2023}, but the expression was proposed \textit{ad hoc}, preventing generalization to other contexts. Thus, we still lack a way to derive diffusion coefficients of bacteria and other active swimmers in disordered porous media from first principles.

Here, we study diffusion of a run-and-tumble particle navigating a 2D disordered porous environment consisting of hard, overlapping circles. To approach the problem analytically, we abstract away the environment, and instead consider states defined by the number of obstacles the cell is in contact with. Then, we approximate transitions between these states as being Markovian and derive the transition rates using knowledge of the environment statistics and cell-surface interactions. This allows us to derive the effective diffusion coefficient in terms of cell and environmental parameters, which shows excellent agreement with simulations with explicit obstacles. The diffusivity depends non-monotonically on run length \cite{licata_diffusion_2016, volpe_topography_2017, bertrand_optimized_2018, kurzthaler_geometric_2021}; we solve the optimal run length, finding a new scaling as a function of obstacle density. Finally, the diffusivities for all obstacle densities can be collapsed onto a single universal curve. Our results suggest that a relatively small set of features are sufficient to describe macroscopic diffusion of bacteria in porous media.



\begin{figure}[t]
\includegraphics[]{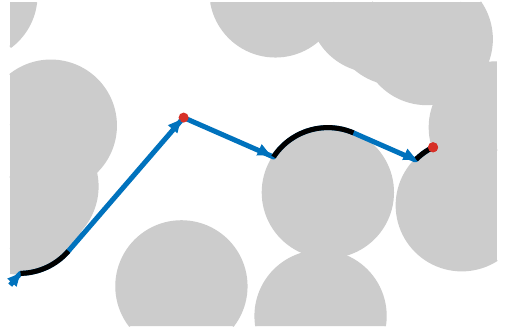}
\caption{Model. A run-and-tumble particle navigates a model porous medium. The obstacles are hard overlapping circles (gray). The cell swims straight (blue) until it tumbles (red). When in contact with an obstacle (black), the component of its velocity projected onto the obstacle surface normal is zero.}
\label{fig:schematic}
\end{figure}

\textit{Results ---} To model a bacterium navigating through a porous medium, we consider a run-and-tumble particle in a 2D environment (Fig. \ref{fig:schematic}). To create pore structures, we use the Lorentz gas model \cite{van_beijeren_transport_1982, machta_diffusion_1983, zeitz_active_2017}, a prototypical model for studying transport in porous media \cite{torquato_effect_2012}. In this model, hard circles are placed independently at random and can overlap, and the number of obstacles in a given area is Poisson-distributed. 

In free space, the cell swims straight with speed $v$ in direction $\uhat(t) = (\cos{\phi(t)}, \ \sin{\phi(t)})$, where $\phi(t)\in[0,2\pi]$ is the angle between the cell's heading and the $x$-axis. The cell tumbles at rate $\lambda$ \cite{berg_chemotaxis_1972, berg_e_2004}, causing it to instantaneously randomize its heading. For simplicity, we exclude the effects of hydrodynamics, which can suppress tumbles near surfaces \cite{molaei_failed_2014}. We also exclude other flagellar dynamics, such as reversal of the bundle when the cell encounters a surface \cite{cisneros_reversal_2006, wu_swarming_2020}. When in contact with an obstacle, cell position follows \cite{saintillan_dispersion_2023}: 
\begin{equation}\label{eq:dyn}
\begin{split}
    \dot{\xm} &= v \left( \uhat(t) - \alpha(t) \ \bm{n}(\xm,t) \right),
\end{split}
\end{equation}
\noindent where $\bm{n}(\xm,t))$ is the surface normal vector at the cell's position $\xm(t)$ and $\alpha(t) = \uhat(t) \cdot \bm{n}(\xm,t)$.

To non-dimensionalize the problem, we rescale length by the obstacle radius $R$ and time by $R/v$. This leaves two dimensionless parameters: the cell's mean run length $\beta=v/(\lambda R)=L_r/R$, and the mean distance between obstacles $\gamma=L_c/R$. In this environment, the mean chord length is $\gamma = 1/(2 \rho)$, where $\rho$ is the dimensionless number density of obstacles. Furthermore, $\gamma$ fully determines the fraction of void space, $\phiv = \exp(-\pi/(2 \gamma))$ \cite{strieder_variational_1973,doi_new_1976}. Finally, the effective diffusion coefficient in free space is $\Dliq = \beta/d$, where $d=2$ is the number of spatial dimensions.


To derive cells' long-time behavior, it is often useful to consider a Fokker-Planck equation for cell density that is equivalent to the dynamics in Eqn. \ref{eq:dyn}. But this ``microscopic" description is complicated, even when the walls have simple shapes and known locations \cite{ezhilan_distribution_2015, elgeti_run-and-tumble_2015, alonso-matilla_transport_2019, chen_shape_2021}. Instead, we write down a different microscopic model, which is wrong, but that aims to keep the features that matter at ``macrosopic" length and time scales. 

To this end, we abstract the dynamics to three states: state 0, when the cell is in the void space; state 1, when the cell is contact with one obstacle; and state 2, when the cell is in contact with two obstacle. Then, we make a dramatic simplification by approximating the transitions among these states as being Markovian.

First, we need to model the motion of the cell in each state. In state 0, the cell moves freely with heading $\uhat$ and dimensionless speed 1. In state 1, since displacements perpendicular to the cell's heading have a small effect on the diffusivity \cite{saintillan_dispersion_2023}, we model cells as moving in direction $\uhat$ but with reduced speed $\nu \approx 1/2$ (SI). Finally, cells in contact with two obstacles cannot move, so state 2 is the ``trapped" state observed in experiments, while states 0 and 1 together constitute the ``hopping" state \cite{bhattacharjee_bacterial_2019,bhattacharjee_confinement_2019}.

\begin{figure}[t]
	\includegraphics[]{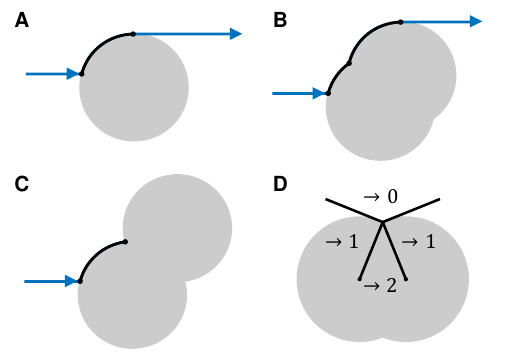}
	\caption{Transitions among states. A) A cell in free space (state 0) can encounter an obstacle, transitioning to state 1. It can transition back to state 0 by sliding off the obstacle or tumbling. B-C) A cell in contact with one obstacle (state 1) can encounter a second obstacle and either: (B) continue sliding and remain in state 1; or (C) become trapped and transition to state 2. D) From state 2, the cell can escape to state 0 or 1 by tumbling. For this configuration, post-tumble headings that fall in each sector transition the cell to the state indicated.}
	\label{fig:transitions}
\end{figure}

Next, we need to enumerate the transitions among these states and derive the transition rates (SI). From state 0, the cell can only transition to state 1 by encountering an obstacle (Fig. \ref{fig:transitions}A), with rate $k_{01}=\gamma^{-1}$. In state 1, the cell can slide off the obstacle at rate $k_{10}\approx2/\pi$, or it can encounter a second obstacle. The second encounter can lead to one of two outcomes: either the cell continues sliding on the second obstacle, thus staying in state 1 (Fig. \ref{fig:transitions}B), or it gets trapped at the intersection point and transitions to state 2 (Fig. \ref{fig:transitions}C). Assuming that nearly all encounters with a second obstacle lead to trapping, the state 1 to 2 transition occurs with rate $k_{12} \approx 2/\pi \ \gamma^{-1}$. Transitions due to encountering or sliding off an obstacle maintain heading $\uhat$. Finally, tumbles occur with rate $\beta^{-1}$ and can cause state transitions (Fig. \ref{fig:transitions}D). The probability that a tumble in state $i$ causes a transition to state $j$ is denoted $p_{ij}$ (SI): $p_{10}=1-p_{11}=1/2$, $p_{20}\approx1/4$, $p_{21}=1/2$, and  $p_{22}=1-p_{20}-p_{21}$.

With cell motion in each state and the transition rates, we arrive at evolution equations for the probability density $\psi_i(\xm,t,\uhat)$ of positions $\xm$ and headings $\uhat$ in state $i$: 
\begin{widetext}
\begin{equation}\label{eq:3pdes}
\begin{split}
\partial_t \psi_0(\xm,t,\uhat) + \nabla_x \cdot (\uhat \ \psi_0) &= 
-\left( \beta^{-1} + k_{01} \right) \psi_0 + k_{10} \ \psi_1 + \frac{\beta^{-1}}{2\pi} ( P_0 + p_{10} \ P_1 + p_{20} \ P_2 ) \\
\partial_t \psi_1(\xm,t,\uhat) + \nu \ \nabla_x \cdot (\uhat \ \psi_1) &= 
-\left(\beta^{-1} + k_{10} + k_{12} \right) \psi_1 + k_{01} \ \psi_0 + \frac{\beta^{-1}}{2\pi} \left( p_{11} \ P_1 + p_{21} \ P_2 \right)\\
\partial_t \psi_2(\xm,t,\uhat) &= 
-\beta^{-1} \ \psi_2 + k_{12} \ \psi_1  + \frac{\beta^{-1}}{2\pi} \ p_{22} \ P_2
\end{split}
\end{equation}
\end{widetext}
\noindent Here, $P_i(\xm,t) = \int \psi_i(\xm,t,\uhat) \ \dd \uhat$ is the density of cells in state $i$ and position $\xm$, marginalized over headings $\uhat$. The terms on the right-hand side of each equation describe the transitions among states. 

By eliminating the obstacles, the model lost the fact that obstacles exclude volume. For a large control volume, some fraction of its surface is blocked by obstacles and does not allow flux through. In 2D isotropic media, the fraction that is unblocked is exactly the void space volume fraction $\phiv$, and therefore all fluxes must be proportional to $\phiv$. Indeed, for passive diffusion, $\Deff$ is upper bounded by $\Dliq \ \phiv$ \cite{strieder_variational_1973}. To correct for the excluded volume of the obstacles, we will need to introduce a factor of $\phiv$ into the diffusion coefficient derived from Eqn. \ref{eq:3pdes}.


The model in Eqn. \ref{eq:3pdes} is an intermediary towards deriving a diffusion coefficient. Standard techniques of homogenization can be used to derive the long-time diffusive behavior of models like Eqn. \ref{eq:3pdes} \cite{sagues_diffusive_1986,mccarty_effective_1988,erban_individual_2004,xue_multiscale_2009,bensoussan_asymptotic_2011} (SI). First, we introduce a small parameter $\eps \ll 1$. For diffusion, we expect typical displacements to scale like the square root of time, $\Delta x \sim \Delta t^{1/2}$, and so we change variables to long length and time scales $\xi = \eps \ x$ and $\tau = \eps^2 \ t$. Finally, we expand the probability density of swimmers in each state $\psi_i(\xm,t,\uhat)$ in powers of $\eps$, plug the expansion into Eqn. \ref{eq:3pdes}, and collect terms by order of $\eps$. The $O(\eps^2)$ equation will ultimately become an effective diffusion equation:
\begin{equation}\label{eq:diffeq}
\partial_t \Ptot(\xim,\tau) = -\nabla_\xi \cdot (\J^{(1)}_0 + \nu \ \J^{(1)}_1).
\end{equation}
\noindent Superscripts with parentheses indicate the order of $\epsilon$ in the asymptotic expansion, while subscripts indicate the state. $\Ptot(\xim,\tau) = P_0^{(0)}+ P_1^{(0)} +P_2^{(0)}$ is the zeroth-order density of cells, summed over states; $P_i^{(0)} = \int \psi_i^{(0)}(\xm,t,\uhat) \ \dd \uhat$ is the zeroth-order density of cells in state $i$; and $\J_i^{(1)}(\xim,\tau) = \int \uhat \ \psi_i^{(1)}(\xm,t,\uhat) \ \dd \uhat$ is the first-order flux of cells in state $i$. 

To derive the diffusion coefficient, we need to solve the first-order fluxes, which in turn depend on the solutions to the zeroth-order equations. At zeroth order, there are no derivatives, so we get a system of algebraic equations for $p_i$, the steady state fraction of time spent in state $i$:
\begin{equation}\label{eq:p_i}
\begin{split}
    p_0 = \frac{1}{1+a+b}, \ p_1 = \frac{a}{1+a+b}, \ p_2 = \frac{b}{1+a+b} \\
    a = \frac{k_{01}}{\beta^{-1} \ p_{10} + k_{10} + k_{12} \frac{p_{20}}{p_{20}+p_{21}}}, \ b = a \ \frac{\beta \ k_{12}}{(1-p_{22})}
\end{split}
\end{equation}
Then, solving the first order fluxes $\J_i^{(1)}$, we find that they are proportional to the gradient of total cell density $\nabla_\xi \Ptot$, as expected for diffusion. Together with Eqn. \ref{eq:diffeq} and a factor of $\phiv$, we find the effective diffusion coefficient, $\Deff$:
\begin{equation}\label{eq:Deff}
\begin{split}
\Deff & = \frac{\phiv}{d} \Bigg( \frac{1}{\Lambda_0} p_0 + \frac{\nu}{\Lambda_0} \ \frac{k_{10}}{\Lambda_1} \ p_1  \\ & + \frac{\nu}{\Lambda_1} \ \frac{k_{01}}{\Lambda_0} \left( p_0 +\nu \ \frac{k_{10}}{\Lambda_1} \ p_1 \right) + \frac{\nu^2}{\Lambda_1} \ p_1 \Bigg).
\end{split}
\end{equation}
$\Deff$ has contributions from flux in state 0 (first term), flux in state 1 (last term), and fluxes that are preserved when the cell changes state (second and third terms).  $\Lambda_1 = \beta ^{-1} + k_{10} + k_{12}$ is the rate at which flux in state 1 becomes decorrelated due to tumbles, sliding off an obstacle, or encountering a second obstacle, respectively. $\Lambda_0 = \beta^{-1} + k_{01}\ (1-k_{10}/\Lambda_1)$ is the decorrelation rate of flux in state 0, due to tumbles and encounters with obstacles. The latter is suppressed because, with probability $k_{10}/\Lambda_1$, the cell can re-enter state 0 by sliding off an obstacle before tumbling, which maintains its heading.


To test this theory, we simulated run-and-tumble point particles moving through many realizations of 2D environments with explicit circular obstacles (SI). Since obstacle positions are independent of each other, the simulation built the environment as the particle explored it, reducing computation time. Cell motion while sliding on an obstacle was computed analytically, as were the times of encounters and slide-off events. We only simulated environments with $\gamma > \gamma_c^{2D} \approx 1.39$ \cite{bauer_localization_2010}, where $\gamma_c^{2D}$ is the void space percolation threshold. Below this threshold, the void spaces are highly likely to form closed pores, and there is no diffusive behavior.

\begin{figure}[t]
\includegraphics[]{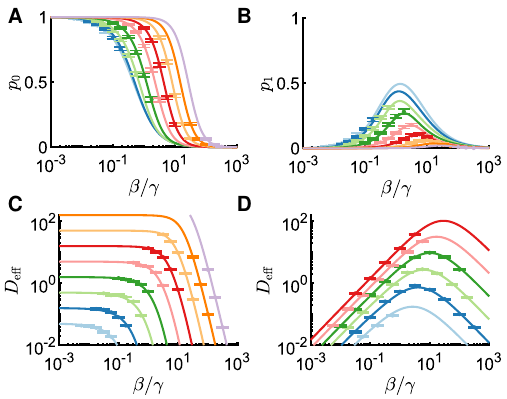}
\caption{Predictions and simulations. A-B) Fraction of time spent in state 0, $p_0$, (A) and state 1, $p_1$ (B). Lines are predictions of Eqn. \ref{eq:p_i} and dots are simulations. C-D) Effective diffusion coefficients. Lines are predictions of Eqn. \ref{eq:Deff} and dots are simulations. In (A-C), colors are fixed values of $\beta \in [10^{-1}, 10^3]$, increasing by factors of $10^{1/2}$ from light blue to light purple, and varying $\gamma$. In (D), colors are fixed $\gamma \in [10^{1/4}, 10^{3/2}]$, increasing by factors of $10^{1/4}$ from light blue to red, and varying $\beta$.}
\label{fig:predictions}
\end{figure}

The theoretical predictions for the fraction of time in each state $p_i$ (Eqn. \ref{eq:p_i}; Fig \ref{fig:predictions}AB) and the diffusion coefficient $\Deff$ (Eqn. \ref{eq:Deff}; Fig \ref{fig:predictions}CD) agree very well with simulations. Similar to past studies \cite{licata_diffusion_2016, volpe_topography_2017, bertrand_optimized_2018, kurzthaler_geometric_2021}, Fig. \ref{fig:predictions}D shows that $\Deff$ depends non-monotonically on the mean run length $\beta$. In particular, it increases proportionally to $\beta$ for small $\beta$ and decreases proportionally to $\beta^{-1}$ for large $\beta$. Eqn. \ref{eq:Deff} accurately captures the heights and locations of these peaks.

We can find approximate expressions for the optimal run length and maximum diffusivity in the regime of dilute obstacles, $\gamma \gg 1$ (SI). In this limit, sliding becomes fast compared to obstacle encounters, and a cell in state 1 rapidly slides off or becomes trapped. Thus, state 1 disappears, and the cell effectively switches between swimming freely and being trapped (SI):
\begin{equation}\label{eq:Ddilute}
    \Deff \approx \frac{\phiv}{d} \frac{p_0}{\Lambda_0} = \frac{\phiv}{d} \frac{1}{\beta^{-1}+\ktrap} \frac{\kesc}{\kesc+\ktrap}.
\end{equation}
Here, $\ktrap= k_{01} k_{12} / k_{10} \propto \gamma^{-2}$ is the effective trap rate, which is balanced by the escape rate, $\kesc = \beta^{-1} \ (1-p_{22})$, to determine the fraction of time swimming, $p_0$. In the SI, we also derive $\Deff$ to sub-leading order in $\gamma$.

Maximizing the dilute-obstacle expression for $\Deff$ in Eqn. \ref{eq:Ddilute} with respect to $\beta$ (SI), we find: 
\begin{equation}\label{eq:betaopt}
\begin{split}
\beta^* &= \ktrap^{-1}\ \sqrt{1-p_{22}}  + \mathcal{O}(\gamma)
\\ 
\Deff^* &= \frac{\phiv}{d} \frac{\beta^*}{2+c} + \mathcal{O}(\gamma)\\
c &= \frac{2-p_{22}}{(1-p_{22})^{1/2}} + \mathcal{O}(\gamma^{-1})
\end{split}
\end{equation}
\noindent The optimal run length here scales like $\beta^* \sim \gamma^{2}$, because trapping requires two obstacle encounters, and encounters are rate-limiting. As in other works, to maximize diffusion, the cell should make its runs comparable to the distance between traps, but that distance is modulated by how costly it is to get trapped ($p_{22}$). 

\begin{figure}[b]
\includegraphics[]{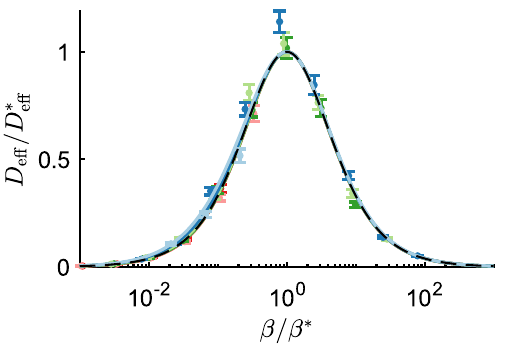}
\caption{Diffusivities collapse. Rescaling the diffusion coefficient by its maximum, $\Deff/\Deff^*$,  and plotting against run length rescaled by the optimal value, $\beta/\beta^*$, causes all diffusivities to collapse onto a universal curve, independent of $\gamma$. Lines and dots are the same as in Fig. \ref{fig:predictions}D. Each $\Deff^*$ and $\beta^*$ was found numerically here. The black dashed line is the universal curve in Eqn. \ref{eq:universal}, with no free parameters.
}
\label{fig:collapse}
\end{figure}

Finally, plotting $\Deff/\Deff^*$ against $\delta = \beta/\beta^*$, we find that all of the theoretical curves and simulation data collapse onto a single, universal curve (Fig. \ref{fig:collapse}):
\begin{equation}\label{eq:universal}
\frac{\Deff(\delta)}{\Deff^*} \approx \frac{(2+c) \ \delta}{1+c \ \delta+\delta^2},
\end{equation}
\noindent where $c$ is given in Eqn. \ref{eq:betaopt}. This expression shows good agreement with the rescaled theory and simulations, even outside the dilute-obstacle regime. In the dilute limit, this collapse occurs because the obstacle radius $R$ becomes small. Since the problem effectively loses a length scale, it also loses a dimensionless parameter, resulting in the curve collapse. The relevant time scale becomes the average time between trap events $(R/v) \ \ktrap^{-1}$, and the relevant length scale is the average distance traveled between trap events $R\ \ktrap^{-1}$ (SI). 

%

\textit{Discussion ---} Here, we derived an analytical expression for the diffusion coefficient of a run-and-tumble bacterium navigating through a 2D, disordered porous environment. Our approach was to ``forget" 
the environment geometry: instead, we considered cell states, defined by how many obstacles the cell contacts, and assumed Markovian transitions among them. This simplified model allowed us to derive a diffusion coefficient that shows very good agreement with simulations in explicit geometries. 

In the regime of dilute obstacles, we found that the optimal run length scales like the square of the mean chord length, $\beta^* \sim \gamma^2$ (Eqn. \ref{eq:betaopt}). This is different from the $\beta^* \sim \gamma$ scaling, or similar, observed in other works \cite{licata_diffusion_2016, bertrand_optimized_2018, kurzthaler_geometric_2021}, and it arises from the fact that the cell must encounter two obstacles before trapping occurs. In 3D, we expect $\beta^* \sim \gamma^3$ for $\gamma\gg1$ because three obstacles must be encountered for trapping. However, unlike in 2D, diffusive behavior is possible for $\gamma<1$ in 3D, since the void percolation threshold is $\gamma_c^{3D}\approx 0.379$ \cite{hofling_localization_2006}. Therefore, $\beta^* \sim \gamma$ scaling is possible for small $\gamma$ in 3D. If the obstacle surfaces are concave, such as when the void spaces are inside the circles \cite{volpe_topography_2017, kurzthaler_geometric_2021}, trapping occurs upon the first contact with an obstacle. This likely leads to $\beta^* \sim \gamma$ scaling of the optimal run length, in both 2D and 3D. 

We also found that all of the diffusion coefficients collapsed onto a single curve when rescaled appropriately, and we derived an analytical expression for this universal curve (Eqn. \ref{eq:universal}). A major implication of this collapse is that the diffusion coefficient can be determined from a small number of quantities. In the dilute obstacle regime (Eqn. \ref{eq:Ddilute}), these are just the run length $\beta$, the mean ``hop" time, $\ktrap^{-1}$, and the mean ``trap" time, $\kesc^{-1}$, all of which are accessible experimentally. Furthermore, we expect bacteria with different swimming patterns to fall on the same universal curve, because it depends weakly on the probability that reorientation leads to escape, $p_{22}$. There is already some evidence for this collapse in simulations by others \cite{kurzthaler_geometric_2021}. Finally, diffusivities in 3D should also collapse when $\gamma\gg1$, because the obstacle radius becomes an irrelevant length scale. 

Using this model, we can ask what might set the distributions of hop and trap durations observed in experiments \cite{bhattacharjee_bacterial_2019,bhattacharjee_confinement_2019}. Because of our assumption of Markovian state transitions, all wait-time distributions predicted by Eqn. \ref{eq:3pdes} are sums of exponentials. Since hopping is the union of states 0 and 1, our model predicts that the hop time distribution is a sum of two exponentials, which we derive in the SI. 
Additionally, experiments found that trap durations were power-law distributed, with an exponent between -2 and -3 \cite{bhattacharjee_bacterial_2019,bhattacharjee_confinement_2019}. One way this can emerge is by mixing exponentials with decay rates that vary over orders of magnitude \cite{tu_how_2005,costa_fluctuating_2023}. Here, although the escape rates depend on the configuration of the obstacles that trap the cell, they only vary by a factor of 4 (SI), which is not enough to generate a power law over multiple decades. However, this variation in escape rates can act in concert with other sources of variation to produce a power law. Steric interactions, which can prevent a non-spherical cell from reorienting, might broaden the distribution of escape rates. Additionally, there is always cell-to-cell and within-cell variability in tumble rates \cite{spudich_non-genetic_1976, korobkova_molecular_2004, park_interdependence_2010, masson_noninvasive_2012, dufour_direct_2016, waite_non-genetic_2016, waite_behavioral_2018, mattingly_escherichia_2021, mattingly_collective_2022}, which could also contribute to the observed power-law distribution of trap times. 

Our theoretical approach can be used to make progress on several related problems. Diffusion past non-overlapping spherical obstacles may be possible to derive, using statistics of that environment \cite{lu_lineal-path_1992,lu_chordlength_1993,torquato_chord-length_1993}. Rotational diffusion can be included as another way for cells to escape from surface states. A 3D environment can be treated in nearly the same way, but with an additional surface state and different transition rates. For non-spherical obstacles, additional states are needed if the cell's motion is different on different obstacle faces. In general, the states, the possible transitions, and the transition rates depend on the shapes of the obstacles, the statistics of their arrangements, and on the cell's motion when in contact with obstacles. Finally, bacterial chemotaxis in shallow gradients can readily be modeled this way by including internal states coupled to an external chemical field \cite{erban_individual_2004, xue_multiscale_2009, celani_bacterial_2010, si_pathway-based_2012, reichhardt_active_2014, grognot_physiological_2023}. Several scenarios are much more difficult to study using our approach. Non-circular cells are a challenge because the steric interactions with obstacles complicate reorientation dynamics \cite{chen_shape_2021,kurzthaler_geometric_2021}. Hydrodynamic interactions and fluid flows are also difficult to include in this model \cite{dehkharghani_bacterial_2019, de_anna_chemotaxis_2020, scheidweiler_trait-specific_2020, dentz_dispersion_2022}.

In all, we present a promising approach for deriving diffusion coefficients of active microswimmers in disordered porous media from first principles.

\begin{acknowledgments}
{\it Acknowledgments ---} The author thanks Mike Shelley and the Biophysical Modeling Group at Flatiron for helpful feedback---in particular, Brato Chakrabarti, Scott Weady, Bryce Palmer, Ido Lavi, Suryanarayana Maddu, Victor Chard\`es, and Vicente G\'omez Herrera. Thanks also to David Saintillan for helpful discussions. Colors in Figs. \ref{fig:predictions} and \ref{fig:collapse} are those of ref \cite{harrower_colorbrewerorg_2003}, as implemented in Matlab by ref \cite{stephen23_colorbrewer_2023}. This work was supported by the Simons Foundation.
\end{acknowledgments}

\bibliography{refs_bibtex}

\end{document}